\documentclass[aps,prd,twocolumn,noeprint,showpacs,nofootinbib,superscriptaddress]{revtex4-2}
\usepackage{calligra}
\DeclareMathAlphabet{\mathcalligra}{T1}{calligra}{m}{n}
\DeclareFontShape{T1}{calligra}{m}{n}{<->s*[2.2]callig15}{}
\newcommand{\scriptr}{\mathcalligra{r}\,}

\usepackage{graphicx}
\usepackage{dcolumn}
\usepackage{bm}

\usepackage{graphicx}
\usepackage{color}
\usepackage{natbib}
\usepackage{epsfig}
\usepackage{float}
\usepackage{xcolor}
\usepackage{booktabs}
\usepackage{bibunits}

\usepackage{amsmath,amssymb,amsfonts}

\begin{document}

\preprint{APS/123-QED}

\title{A neural-network-based surrogate model for the properties of neutron stars in 4D Einstein-Gauss-Bonnet gravity}

\author{Ioannis Liodis}
\affiliation{Department of Physics, Aristotle University of Thessaloniki, 54124 Thessaloniki, Greece}

\author{Evaggelos Smirniotis}
\affiliation{Department of Physics, Aristotle University of Thessaloniki, 54124 Thessaloniki, Greece}

\author{Nikolaos Stergioulas}
\affiliation{Department of Physics, Aristotle University of Thessaloniki, 54124 Thessaloniki, Greece}

\date{\today}

\begin{abstract}

Machine learning and artificial neural networks (ANNs) have increasingly become integral to data analysis research in astrophysics due to the growing demand for fast calculations resulting from the abundance of observational data. 
Simultaneously, neutron stars and black holes have been extensively examined within modified theories of gravity since they enable the exploration of the strong field regime of gravity. In this study, we employ ANNs to develop a surrogate model for a numerical iterative method to solve the structure equations of NSs within a specific 4D Einstein-Gauss-Bonnet gravity framework. We have trained highly accurate surrogate models, each corresponding to one of twenty realistic EoSs. The resulting ANN models predict the mass and radius of individual NS models between 10 and 100 times faster than the numerical solver. In the case of batch processing, we demonstrated that the speed up is several orders of magnitude higher. We have trained additional models where the radius is predicted for specific masses. Here, the speed up is considerably higher since the original numerical code that constructs the equilibrium models would have to do additional iterations to find a model with a specific mass. Our ANN models can be used to speed up Bayesian inference, where the mass and radius of equilibrium models in this theory of gravity are required.

\end{abstract}

\maketitle

\section{Introduction}
    \label{sec:intro}
    Recent years have witnessed an abundance of observational data on NSs coming both from gravitational wave detectors, namely Advanced LIGO \cite{LIGO_1} and Advanced Virgo \cite{Virgo_1}, and electromagnetic radiation, like the NICER mission \cite{2014SPIE.9144E..20A}. These astrophysical observations have facilitated a significant number of attempts to constrain the EoS of NSs, including the NICER mass and radius measurements \cite{Riley_2019,Miller_2019,Miller_2021}, the tidal deformability measurement through GWs \cite{Van_Oeveren_2017,Hinderer:2007mb,Binnington:2009bb,Damour:2009vw,Chatziioannou2020,Dietrich2021}, as well as joint constraints, e.g., \cite{Raaijmakers_2020,PhysRevD.106.043012,Biswas_2021,Biswas_2022,Traversi_2020,Xie_2019,PhysRevD.103.103015,Tim_Dietrich,Landry_2020PhRvD.101l3007L,Raaijmakers:2021uju}. Especially the binary NS merger detection GW170817 \cite{LIGOScientific:2017vwq,LIGOScientific:2018hze} has sparked further studies in this field \cite{GW170817_Jiang_2019,GW170817.120.172702,GW170817.120.261103,LIGOScientific:2018cki,GW170817_Landry_2018,GW170817.120.172703,GW170817.121.062701,GW170817.99.123026}.
    
    Several studies have employed Artificial Neural Networks (ANNs) to reconstruct the EoS of NSs based on their observable properties \cite{PhysRevD.101.054016,PhysRevD.98.023019,Fujimoto_2021,Ferreira_2021,galaxies10010016}. For instance, \cite{morawski_2020} investigated the use of ANNs supported by the autoencoder architecture, while \cite{Soma_2022,Soma_2023} employed ANNs to represent the EoS in a model-independent way, exploiting the unsupervised automatic differentiation framework. Additional machine-learning techniques have been applied to investigate the NS EoS. For example, in \cite{Lobato_2022},  a clustering method was utilized to identify patterns in mass-radius curves, and \cite{lobato2022unsupervised} explored correlations among different EoSs of dense matter using unsupervised Machine Learning (ML) techniques. Furthermore, attempts have been made to derive nuclear matter characteristics from NS EoS and observations using deep neural networks, see, e.g., \cite{Ferreira_2022,krastev2023deep}.

    NSs have also been the subject of theoretical investigations within alternative theories of gravity - see \cite{2022arXiv221101766D,Berti_2015} for comprehensive reviews and possible tests and \cite{Charmousis_2022} for the particular theory considered in this work. In particular, the high mass of the secondary component in the GW190814 event \cite{LIGOScientific:2020zkf} has attracted significant interest, as it can only be explained in General Relativity (GR) under extreme conditions. Specifically, the only possibilities of this object being a NS is either having an extremely stiff EoS or being the fastest rotating NS ever observed \cite{Huang_2020,Bombaci_2021,Roupas_2021,Zhou_2021,Zhang_2020,10.1093/mnras/stab1383}. Alternative approaches based on modified gravity theories have been proposed to address this issue as well \cite{Astashenok_2020,Astashenok_2021}. Namely, \cite{Nunes_2020} explored the issue of the maximum mass of NSs, taking into account the thin-shell effect (chameleon screening) on the NS mass-radius relation while considering a soft EoS, thereby demonstrating the possibility of attaining large masses and explaining the secondary component of GW190814 using modified gravity. Additionally, studies involving modified gravity and the GW170817 event have been conducted \cite{Lobato_2020,Lobato_2021}.

    Common methods implemented when attempting to infer the NS EoS from observations of their macroscopic properties are based on Bayesian statistics \cite{Raaijmakers_2020,Miller_2021,PhysRevD.106.043012,Biswas_2021,Biswas_2022,Traversi_2020,Xie_2019,PhysRevD.103.103015,PhysRevLett.126.061101,Tim_Dietrich}. The majority of these algorithms demand a TOV solver to run numerous times before obtaining the final posterior distribution of various parameters. If we wish to incorporate modified theories of gravity in these studies, we will need a modified TOV solver, like the one presented in \cite{Charmousis_2022}. However, this specific algorithm is based on an iterative method for solving the system of differential equations, making Bayesian inference computationally expensive. In any case, solving differential equations numerically thousands or millions of times is significantly time-consuming and can even prove impractical. Therefore, it would be extremely useful to find an alternative way to quickly yet accurately predict the macroscopic properties of NSs, given some defining characteristics or other macroscopic properties of each equilibrium model.

    Driven by the aforementioned motivation, this work focuses on implementing ANN regression for two types of functions: $f_1(\rm{EoS}; \alpha, p_c) \rightarrow (M,R)$ and $f_2(\rm{EoS}; \alpha, M) \rightarrow R$. Here, the EoS represents a distinct variable, since each type includes one ANN model for each EoS. $\alpha$ is the coupling constant of the theory, and $p_c$ is the central pressure of the NS. The first type serves as a surrogate model for the numerical iterative method described in \cite{Charmousis_2022}, which provides the mass and radius of NSs for a specific EoS and a given pair of $\alpha$ and $p_c$. The primary objective is to accelerate the process while maintaining strict accuracy boundaries. The second type cannot be obtained directly using the iterative method since $p_c$ ought to be an input. Consequently, implementing a root-finding algorithm would be the only solution, resulting in further time delays. Conversely, training ANNs to predict $R$ based on $(\rm{EoS}; \alpha, M)$ offers a more straightforward approach to handle type $f_2$. To the best of our knowledge, this is the first work that employs ANNs to predict the bulk properties of NSs within a modified theory of gravity.

    This concept of ANN surrogate models can be implemented in any theory as long as there is a corresponding numerical solver to create a data set for regression. However, the modified theory of gravity studied in this work is a particular 4D Horndeski scalar-tensor model originating from higher dimensional Einstein-Gauss-Bonnet gravity. The action under consideration is \cite{Charmousis_2022}
        \begin{equation}
            S = \frac{1}{2\kappa} \int d^4 x \sqrt{-g} \left( R + \alpha \mathcal{L}
         \right) + S_m,
        \end{equation}
        where $ \kappa = \frac{8\pi G}{c^4}$, $S_m$ is the matter Lagrangian and
        \begin{equation}
            \mathcal{L} = \left[ \phi \mathcal{G} + 4 G_{\mu \nu} \nabla^{\mu} \phi \nabla^{\nu} \phi - 4(\nabla \phi)^2 \Box \phi + 2 (\nabla \phi)^4 \right] ,
        \end{equation}
        where $\mathcal{G}$ is the Gauss-Bonnet scalar
        \begin{equation}
            \mathcal{G} = R^2 - 4R_{\mu \nu}R^{\mu \nu} + R_{\mu \nu \rho \sigma}R^{\mu \nu \rho \sigma} .
        \end{equation}
        The scalar is considered dimensionless, leaving the coupling constant $\alpha$ with dimensions of length squared.
        
        In our work, we are interested in non-rotating relativistic stars, and thus, we briefly present the differential equations considered in \cite{Charmousis_2022} in the Komatsu-Eriguchi-Hachisu
        (KEH)/Cook-Shapiro-Teukolsky (CST) numerical scheme. The line element describing the spacetime geometry of a spherically symmetric star in equilibrium using isotropic coordinates is
        \begin{equation}
            ds^2 = -e^{2 \nu} dt^2 + e^{2 \mu} [ d \scriptr ^2 + \scriptr ^2 (d\theta ^2 + \sin^2 \theta d\varphi^2) ] ,
        \end{equation}
        where $\nu (\scriptr)$ and $\mu (\scriptr)$ are metric functions. For the theory under consideration, one can obtain the field equations, which are elliptic equations for the metric functions
        \begin{subequations}
        \begin{eqnarray}
            \nabla^2 \nu = S_{\nu} (\scriptr),\\
            \nabla^2 \mu = S_{\mu} (\scriptr),
        \end{eqnarray}
        \end{subequations}
        where $\nabla^2 = \partial_{\scriptr\scriptr} + \frac{2}{\scriptr} \partial_{\scriptr}$ is the flat-space Laplacian and $S_{\nu}$, $S_{\mu}$ are the source terms of the elliptic equations.
        The Green's function of the three-dimensional Laplacian operator in spherical coordinates is then used to obtain integral equations. {\color{black}For the scalar field, the equation can be cast in the form of a current conservation equation, see Eq. (A.4) in \cite{Charmousis_2022}}.
        
        Finally, one has to add the hydrostatic equilibrium equation 
        \begin{equation}
            \nabla_{\alpha} (H - \ln{u^t}) = 0 ,
        \end{equation}
        where $u^t = e^{-\nu}$ and $H$ is the specific enthalpy
        \begin{equation}
            H(P) = \int_{0}^P \frac{dP'}{\epsilon (P')+P'} .
        \end{equation}
{\color{black}By fixing, e.g., the central pressure of the star and starting with a trial value for the central value of the scalar field, an iterative numerical method relaxes to the equilibrium solution, using appropriate boundary conditions at infinity. }

\section{Data}
    \label{sec:datasets}
    The data sets were generated using the numerical code described in \cite{Charmousis_2022}. This code requires the coupling constant $\alpha \: [\rm{km}^2]$ of the theory and the central pressure $p_c  \: [10^{35} \: \rm{dyn}/\rm{cm}^2]$ of the NS as inputs to compute the corresponding mass $M \: [\rm{M}_\odot]$ and radius $R \: [\rm{km}]$. For each type of function, we created 20 data sets, one for each of the 20 realistic EoSs listed in Table \ref{tab:EoSs}. Finally, to enhance the training, $(M, R, p_c)$ data were logarithmized, and the entire data set was standardized\footnote{Standardization refers to scaling the data to have zero mean and unit variance to improve the training process of the neural network. For the standardization we used \texttt{\url{https://scikit-learn.org/stable/modules/generated/sklearn.preprocessing.StandardScaler.html}}.}.
    
    \begin{figure*}[t!]
        \centering
        \includegraphics[width=0.8\textwidth]{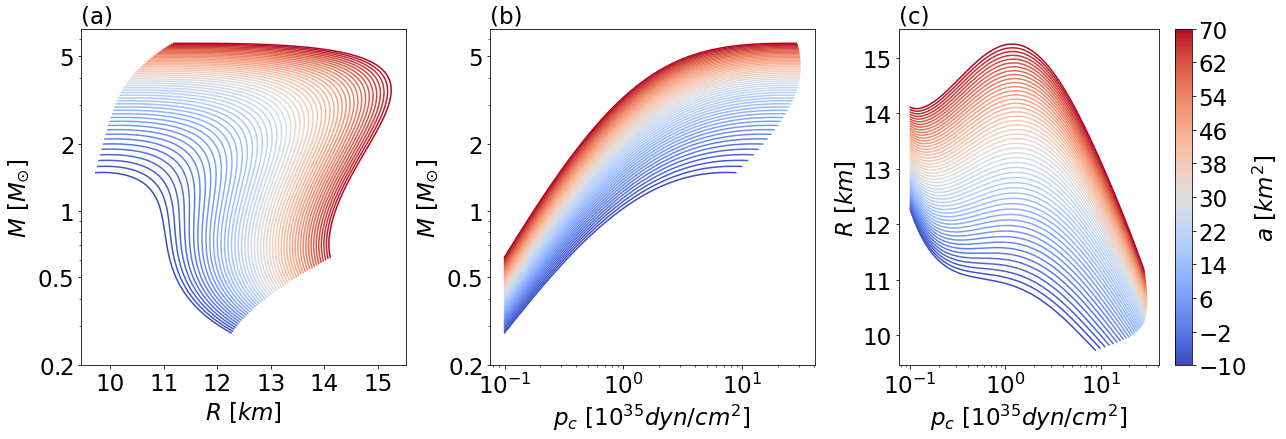}
        \caption{Data set for BSk20, regarding type $f_1$.}
        \label{fig:alldata}
    \end{figure*}
    
    \textit{Type $f_1(\rm{EoS}; \alpha, p_c) \rightarrow (M,R)$.}
    Each data set for this type consists of 51 values of $\alpha \in [-10,70]$ $\times$ 200 values of $p_c \in [0.1, 1.2 \,p_{\rm{max}}]$. The $\alpha$ values were evenly spaced on a linear scale, while the $p_c$ values were logarithmically spaced. Here, $p_{\rm{max}}$ represents the central pressure required for a NS to reach its maximum mass for a specific value of the coupling constant. The upper bound of $p_c$ was chosen to ensure it remains significantly away from the boundary of the stable $M-R$, branch to avoid training difficulties in areas of interest. As a representative example, Figure \ref{fig:alldata} illustrates for EoS BSk20 that each $(\alpha, p_c)$ pair corresponds to a unique $(M,R)$ pair, rendering the training of the ANN relatively straightforward, even when considering a portion of the unstable $M-R$ branch. The same holds for any other EoS in our chosen set.
    
    \textit{Type $f_2(\rm{EoS}; \alpha, M) \rightarrow R$.}
    The data sets for this type have the same size as those for $f_1$, but the range of $p_c$ values was selected differently. The data now consist of 51 values of $\alpha \in [-10,70]$ $\times$ 200 values of $p_c \in [0.1, p_{\rm{max}}]$. This choice is motivated by examining Figure \ref{fig:alldata}. For a specific EoS, the inputs are $\alpha$, which constrains the regression surface to a single curve, and $M$, the remaining variable to determine the output $R$. In case the $M-R$ curve for a specific value of $\alpha$ is non-monotonic, then $f_2$ is not even a function by definition since it would map a single value of its domain (e.g., $M_0$, where $|M_0-M_{\rm{max}}|<<1$) to two different values of its co-domain (e.g., $R_1$, $R_2$). Therefore, the data must be monotonically increasing or decreasing to ensure the function is well-defined.
    
    \begin{table}[]
        \centering
            \caption{Numbering of the EoSs used.}
        \label{tab:EoSs}
        \begin{tabular}{cccr}
        \toprule
            \hspace{0.2cm} Name \hspace{0.2cm} & \hspace{0.2cm}  Number \# \hspace{0.2cm}\\
            \midrule
            APR             &  1 \\
            BHBLP           &  2 \\
            DD2             &  3 \\
            eosAU           &  4 \\
            eosUU           &  5 \\
            BSk20           &  6 \\
            LS220           &  7 \\
            LS375           &  8 \\
            GS1             &  9 \\
            GS2             &  10 \\
            APR3 (PP)       &  11 \\
            ENG (PP)        &  12 \\
            GNH3 (PP)       &  13 \\
            H4 (PP)         &  14 \\
            MPA1 (PP)       &  15 \\
            SLy4 (PP)       &  16 \\
            WFF2 (PP)       &  17 \\
            SFHo            &  18 \\
            TM1             &  19 \\
            TMA             &  20 \\
            \bottomrule
        \end{tabular}
    
    \end{table}

\section{Training and Testing}
    \label{sec:training_testing}
    In this work, the TensorFlow module Keras\footnote{\texttt{\url{https://www.tensorflow.org/api_docs/python/tf/keras}}} was used. For each case, a random selection was made to create a train-test ratio of 70:30. The Mean Square Error (MSE) was chosen as the loss function, while the Absolute Relative Error (ARE) served as the criterion for testing the trained models. A systematic investigation of the network architecture revealed that models with an odd number of dense hidden layers, with a symmetric number of neurons (i.e., 25-35-25) and with alternating two activation functions among the layers (namely ``$\tanh$" and ``$\rm{relu}$") performed better, with lower final loss and Mean ARE (MARE). Based on these findings, the final architecture for each type is shown in Table \ref{tab:architecture}. Initial training attempts involved implementing every optimizer from the list of provided Keras optimizers\footnote{\texttt{\url{https://www.tensorflow.org/api_docs/python/tf/keras/optimizers}}}. However, none of the results were optimal, indicating that the \textit{architecture error}, as defined in \cite{e25010175}, was not the dominant source of error. The \textit{optimization error} was further investigated, leading to the implementation of the second-order {\it Broyden–Fletcher–Goldfarb–Shanno optimizer} algorithm (BFGS), as suggested in \cite{e25010175}. Incorporating BFGS into the training process was not straightforward, as it is not included in the list of provided Keras optimizers. Following a similar approach as described in \cite{BFGS} for BFGS (instead of L-BFGS), {\it the final loss decreased by up to four orders of magnitude} and reached a stable value. The number of total iterations and the final training loss for each EoS are provided in the Appendix, section \ref{sec:iters}.

    \begin{table}[]
        \centering
            \caption{Final network architecture.}
        \label{tab:architecture}
        \begin{tabular}{cccr}
        \toprule
            Layer \hspace{0.2cm} & \hspace{0.2cm} Type $f_1$ \hspace{0.2cm} & \hspace{0.2cm} Type $f_2$ \hspace{0.2cm}\\
            \midrule
            Input layer     &  $(\alpha,p_c)$   & $(\alpha,M)$\\
            Hidden layer 1  &  25-tanh          & 25-tanh\\
            Hidden layer 2  &  35-relu          & 35-relu\\
            Hidden layer 3  &  25-tanh          & 25-tanh\\
            Output layer    &  $(M,R)$          & $R$ \\
            \bottomrule
        \end{tabular}
    \end{table}
    
    \begin{figure*}[t!]
        \centering
        \includegraphics[width=0.8\textwidth]{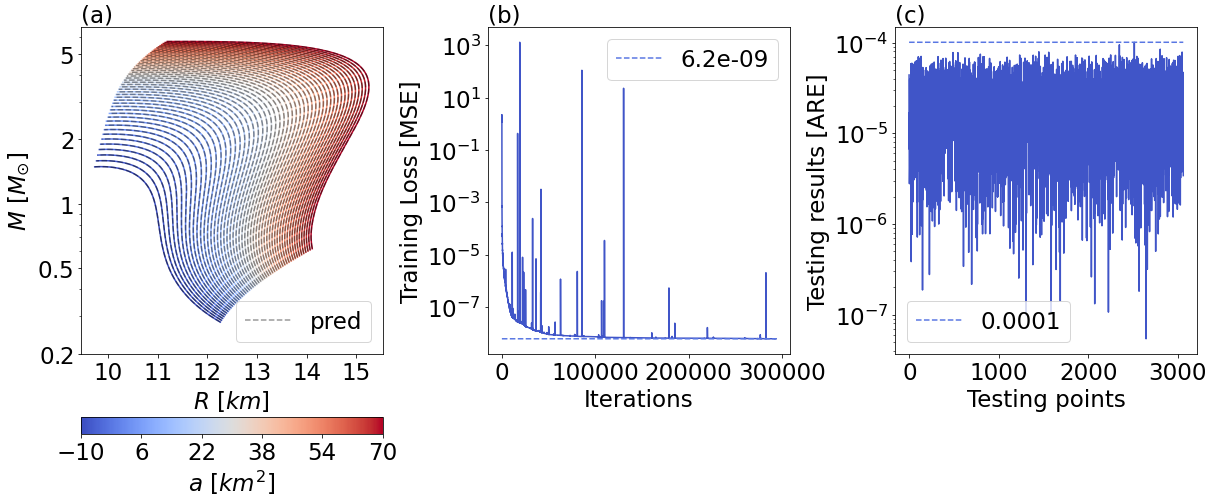}
        \caption{Training results for $f_1$ and BSk20.}
        \label{fig:eos_bsk20}
    \end{figure*}
    
    \begin{figure*}[t!]
        \centering
        \includegraphics[width=0.8\textwidth]{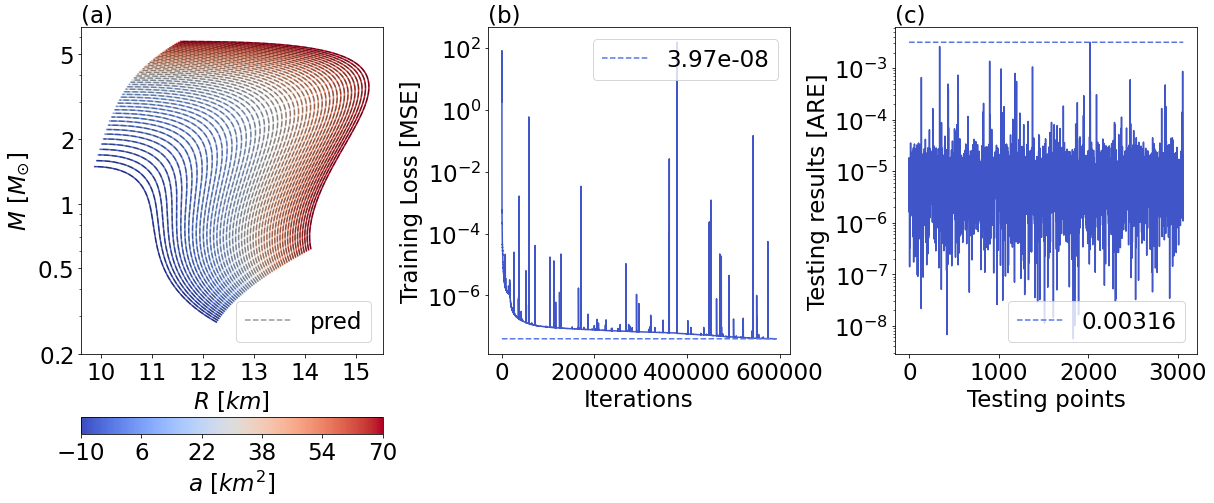}
        \caption{Training results for $f_2$ and BSk20.}
        \label{fig:eos_bsk20_f2}
    \end{figure*}

    \begin{figure*}[t!]
        \centering
        \includegraphics[width=0.8\textwidth]{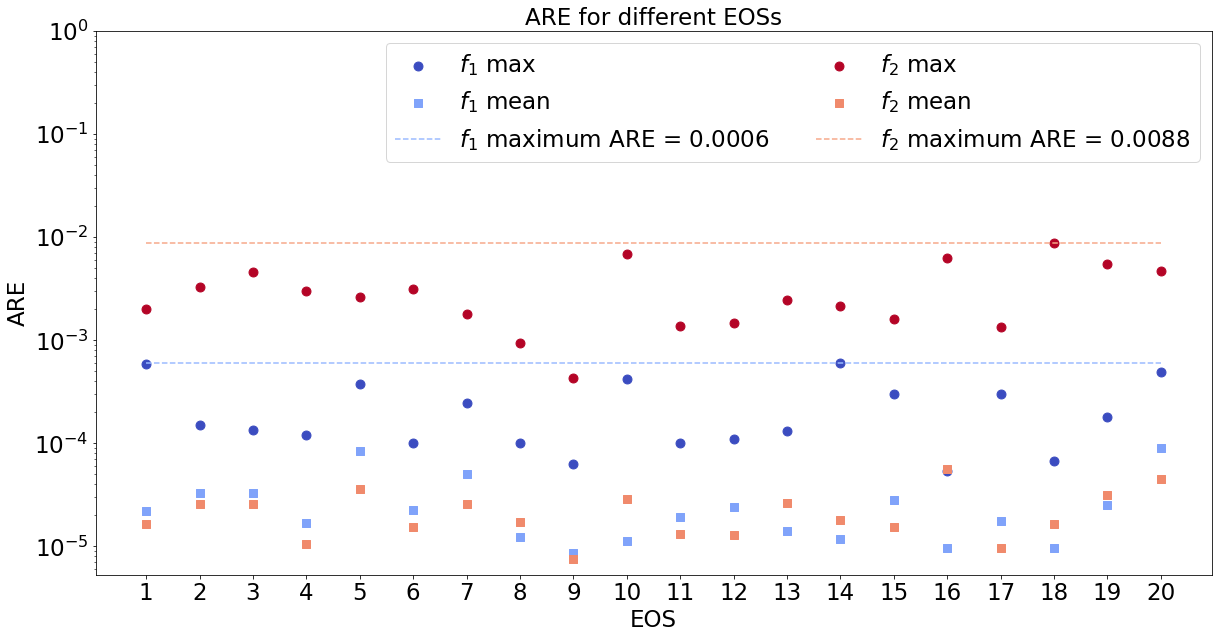}
        \caption{$f_1$ - $f_2$ Absolute Relative Errors on the testing set.}
        \label{fig:f1_f2_ares}
    \end{figure*}

    \begin{figure*}[t!]
        \centering
        \includegraphics[width=0.35\linewidth]{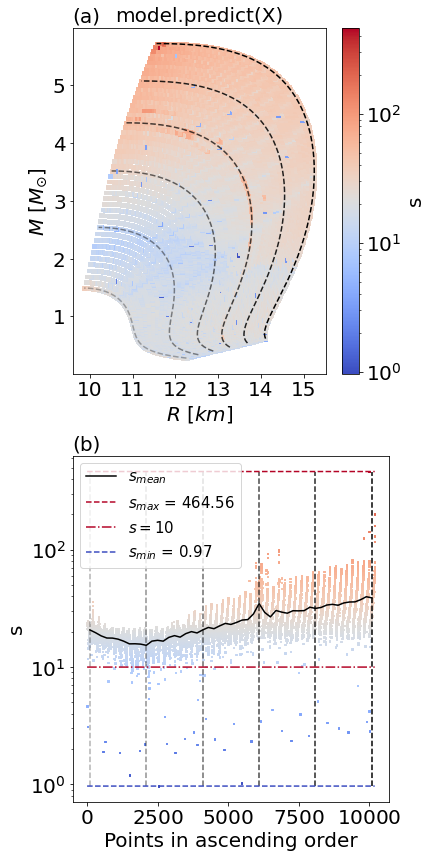}
        \includegraphics[width=0.35\linewidth]{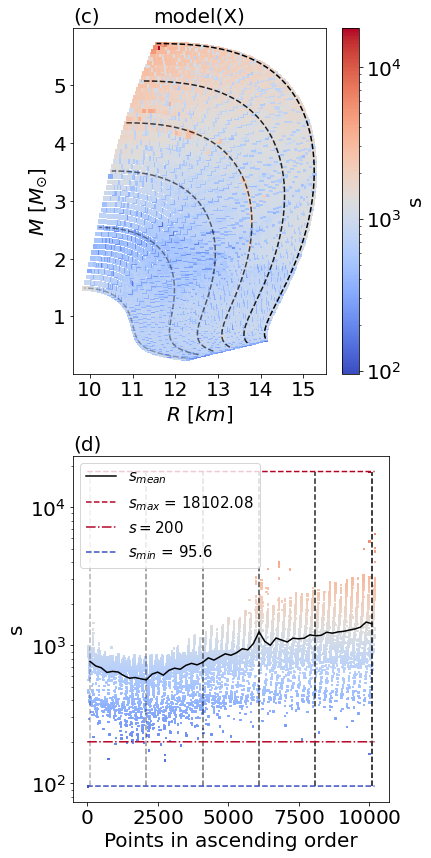}
        \caption{Speed up per $M-R$ data point showing how many times faster the trained ANN model is than the numerical code in each case.}
        \label{fig:f1_ratio}
    \end{figure*}

\section{Results}
    \label{sec:results}
    
    \subsection{Training and Testing results}
    \label{sec:training_res}
    Figures \ref{fig:eos_bsk20} and \ref{fig:eos_bsk20_f2} provide indicative training results for $f_1$ and $f_2$, respectively. More specifically, Figures \ref{fig:eos_bsk20}a and \ref{fig:eos_bsk20_f2}a compare the real to the predicted output. Figures \ref{fig:eos_bsk20}b and \ref{fig:eos_bsk20_f2}b show the training loss per iteration in terms of the Mean Square Error (MSE) and Figures \ref{fig:eos_bsk20}c and \ref{fig:eos_bsk20_f2}c present the Absolute Relative Error (ARE) on the test set. The MSE was chosen as the loss function from Keras' regression loss functions\footnote{\texttt{\url{https://keras.io/api/losses/regression_losses/}}}, utilizing the automatic reduction type. The ARE is defined as
    \begin{equation}
        {\rm ARE}_i = \frac{1}{m} \sum_{j=1}^m \left| \frac{Y_i^j-Y_{i,\rm{true}}^j}{Y_{i,\rm{true}}^j} \right| ,
    \end{equation}
    where $m$ is the number of output neurons, $Y_i^j$ is the {\color{black}network output with index $j$ for test point with index $i$}, and $Y_{i,\rm{true}}^j$ is the corresponding real output. The Mean ARE (MARE) is defined as:
    \begin{equation}
        {\rm MARE} = \frac{1}{n} \sum_{i=1} ^{n} {\rm ARE}_i = \frac{1}{m} \sum_{j=1}^m \left( \frac{1}{n} \sum_{i=1} ^{n} \left| \frac{Y_i^j-Y_{i,\rm{true}}^j}{Y_{i,\rm{true}}^j} \right| \right) ,
    \end{equation}
    where $n=3060$ is the number of points in the test dataset. {\color{black}For $f_1$ models $m = 2$, since there are two output variables ($M$, $R$), while for $f_2$ models $m = 1$ since there is only one output variable ($R$).}
    
    Figure \ref{fig:f1_f2_ares} depicts {\color{black}the MARE and the maximum ARE} on the test set for each EoS, denoted by their corresponding numbers in Table \ref{tab:EoSs}. Both types ($f_1$ and $f_2$) exhibit similar mean behavior on the respective test sets for every EoS, with the MARE ranging between $10^{-5}$ and $10^{-4}$. {\color{black}The maximum ARE is $6\times 10^{-4}$ for $f_1$ and $9\times 10^{-3}$ for $f_2$, demonstrating that in the whole domain of the training and test sets, the absolute relative error never exceeds 1\%. In Appendix \ref{sec:spikes}, we present a more detailed investigation of the distribution of errors in the domain of the training and test sets. }

    \subsection{$f_1$ speed up}
        To assess the speed-up achieved when using the trained ANN models of $f_1$ instead of the numerical code, the dataset corresponding to type $f_2$ was selected. This dataset consists of 51 values of $\alpha \in [-10,70]$ $\times$ 200 values of $p_c \in [0.1, p_{\rm{max}}]$. The choice of this data set for $f_1$ is based on the significance of speed-up in areas of interest, specifically the stable branch of $M-R$ curves. The \textit{speed-up} $s$ is defined as:
        \begin{equation}
            s = \frac{\Delta t_{\rm{ANN}}}{\Delta t_{\rm{num}}},
        \end{equation}
        {\color{black}where $\Delta t_{ANN}$ is the run time when using an ANN model and $\Delta t_{\rm{num}}$ the run time of the numerical code with the iterative numerical scheme.} It is important to note that in our tests, the ANN models ran on an 8-core AMD Ryzen 7 3800X CPU with 16 GB RAM, while the numerical code utilized one such core processor and 8 GB RAM. {\color{black}In principle, running the ANN code on a modern GPU could result in an even higher speed up. The reason we compare the performance on a CPU only is to get an estimate of the speed up in existing applications, such as Bayesian inference, where our ANN model can be integrated without necessarily relying on a GPU.}

        Next, we present three different ways to calculate the output of the models. The speed-up depends on each one of them. It is essential to highlight that by saying \textit{the models}, we refer to the 20 already trained models (one for each EoS), which correspond to $f_1$. The three different ways are:
        \begin{enumerate}
            \item \texttt{model.predict($X$)}, with $X$ being one input value,
            \item \texttt{model($X$)}, with $X$ being one input value,
            \item \texttt{model.predict($\bf X$)}, with $\bf X$ being an array of input values.
        \end{enumerate}

        Since Bayesian inference algorithms mainly require one $(M,R)$ pair per iteration, ways \#1 and \#3 are presented separately, although they use the same function (\textit{predict}) to calculate the outputs. In the next paragraphs, we will present the different speed-ups each of these ways achieves, as well as some other differences they demonstrate.

        \subsubsection{\texttt{model.predict($X$)}}
            The left panel of Figure \ref{fig:f1_ratio} illustrates the speed up for the first case. In panel \ref{fig:f1_ratio}a, the colour represents the speed up, and the dashed lines with increasing transparency correspond to six different $M-R$ curves, for $a = \{-10,6,22,38,54,70\}$ 
            $\rm km^2$, respectively. Panel \ref{fig:f1_ratio}b illustrates the speed-up values for a specific arrangement of the data points. These are arranged in increasing order of $\alpha$, and for each $\alpha$, they are also arranged in increasing order with respect to $p_c$. The color still represents the speed-up values, and each black vertical dashed line corresponds to the same transparency level as in panel  \ref{fig:f1_ratio}a, aiding visualization and establishing the connection between the two plots. {\color{black}Each point on} the solid black line represents the mean value of the speed up $s$ {\color{black}for the 200 points comprising the corresponding $M-R$ curve of a specific value of $\alpha$.} The general trend reveals a minimum on the second vertical dashed line, followed by an increase in performance. It is also evident that the majority of speed-up values range {\color{black}between 10 and 100 (less than 1\% is outside of this range). 
            As a general trend, higher speed-ups are observed for larger input values. }The mean speed up, considering all data points, is {\color{black} $\sim 25$}.

        \subsubsection{\texttt{model($X$)}}
            The right panel of Figure \ref{fig:f1_ratio} illustrates the speed up for the \texttt{model(X)} case, and its configuration is similar to the left panel. It is evident that there is an increase of {\it two orders of magnitude} in speed up. 
            The majority of speed-up values range {\color{black}between 200 and 18000 (less than 0.2\% is outside of this range)}.  The mean speed up, considering all data points, is {\color{black}$> 900$}.
            

        \subsubsection{\texttt{model.predict($\bf X$)}}
            In this case, the $N=51\times200=10200$ data points were inserted as an array of input values $\bf X$. Thus, there is no direct way of comparing its speed to the previous cases. Yet, to give some intuition on the acceleration that this way (of calling the model) provides, we can calculate the \textit{effective run time} $\Delta t_{eff}$, {\color{black}which we define as 
            \begin{equation}
                \Delta t_{eff} = \frac{\Delta t_{N}}{N},
            \end{equation}
where $\Delta t_{N}$  is the {\it total run time} for the whole array as input.} For $N = 10200$ the resulting speed-ups are presented in Table \ref{tab:speed-up}, together with all the previous speed-ups. The first line of the table provides information about the run time of the {\color{black}numerical code that uses the iterative numerical scheme}. The other lines show the corresponding speed-ups. {\color{black}Going from the first to the second and then to the third method, the speed-up increases by 1.5 to 2 orders of magnitude each time. }

{\color{black}Comparing \texttt{model.predict($X$)} to \texttt{model.predict($\bf X$)} it is obvious that the latter does not have a linear behavior regarding the size of $\bf X$}. Thus, we present a detailed investigation of the dependence of the run time of \texttt{model.predict($\bf X$)} on the size of $\bf X$ in Appendix \ref{app:nonlinear}.

            \begin{table}
                \centering
                    \caption{Speed up comparison.}
                    \vspace{0.5cm}
                \label{tab:speed-up}
                \begin{tabular}{ccccr}
                    \toprule
                    Numerical code          &  (Mean)  &(Min)  & (Max) \\
                    run time:          &  1003.5 ms     &  147.96 ms  &  18122.4 ms \\
                    \hline
                    \hline
                    Output \hspace{0.05cm} & \hspace{0.05cm} Speed Up  \hspace{0.05cm} & \hspace{0.05cm} Speed Up \hspace{0.05cm} & \hspace{0.05cm} Speed Up \hspace{0.05cm}\\
                    method \hspace{0.05cm} & \hspace{0.05cm} (Mean) \hspace{0.05cm} & \hspace{0.05cm} (Minimum) \hspace{0.05cm} & \hspace{0.05cm} (Maximum) \hspace{0.05cm}\\
                    \midrule
                    model.predict($X$)        & 25.12        & 0.97     & 464.56      \\
                    model($X$)                & 921.9        & 95.6     & 18102.1     \\
                    model.predict($\bf X$)    & 31295.5      & 4614.5   & 565157.2    \\
                    \bottomrule
                \end{tabular}
            \end{table}

\section{Summary and Discussion}
    {\color{black}The aim of this work was to explore the application of ANNs in predicting the mass-radius relation of NSs for chosen EoSs in a specific alternative theory of gravity}. The data sets used in this study were generated using an {\color{black}iterative numerical code. For each EoS, each equilibrium model is defined by the value of the coupling constant $\alpha$ of the specific} theory of gravity and the central pressure $p_c$ of the NS. Two types of functions, $f_1$ and $f_2$, were considered, each with its own input and output parameters.
    
    For $f_1$, the data sets included 51 values of $\alpha$ ranging from -10 to 70 and 200 values of $p_c$ ranging from $10^{24} \ \rm{\rm{dyn}/\rm{cm}^2}$ to 1.2 times the maximum central pressure ($p_{\rm{max}}$) for each specific EoS. The objective was to predict the NS's mass and radius based on the given values of $\alpha$ and $p_c$. On the other hand, $f_2$ focused on predicting the NS's radius given the $\alpha$ and a fixed value of $M$ for each EoS.
    
    The training and test phase utilized a train-to-test ratio of 70:30, with the mean square error selected as the loss function and the absolute relative error chosen as the testing criterion. An investigation of different ANN architectures revealed that models with an odd number of hidden layers, a symmetric distribution of neurons, and alternating activation functions ``tanh" and ``relu" exhibited lower final loss and mean absolute relative error. The final architecture chosen for training was ``dense 25 tanh - dense 35 relu - dense 25 tanh".
    
    The results demonstrate that the trained ANN models provide accurate predictions for both $f_1$ and $f_2$. The test phase showed that the MARE ranged between $10^{-5}$ and $10^{-4}$, indicating the models' ability to capture the mass-radius relation of NSs across different EoSs.
    {\color{black}The maximum ARE was $6\times 10^{-4}$ for $f_1$ and $9\times 10^{-3}$ for $f_2$, never exceeding 1\% in the whole domain of the training and test sets.}
    
    Furthermore, the $f_1$ speed-up was analyzed to assess the performance improvement achieved using the trained ANN model instead of the iterative numerical code. The results revealed significant acceleration, with most speed-up values {\color{black}(when computing individual models)} ranging from 10 to 100, depending on the input parameters. {\color{black}When the ANN processed a whole array of values, the speed up was several orders of magnitude higher. For $f_2$ the speed up is even higher since the iterative numerical code needs to perform additional iteration to find a model with specific mass.}

    All the constructed models can make accurate predictions in a very short time compared to the iterative numerical solver of differential equations. Therefore, the ANN models can replace the iterative numerical code in Bayesian inference algorithms, paving the way for fast and viable investigations, e.g., of the degeneracy between the uncertainties of the EoS and the theory of gravity.

    It is important to note that further investigations can be pursued to enhance the accuracy and efficiency of the trained ANN models. Future research could explore alternative architectures, optimization algorithms, and additional input parameters to refine the predictions. {\color{black}However, for most practical applications, the accuracy achieved here will be more than sufficient.}
    
    In conclusion, this study successfully employed ANNs to predict the mass-radius relation of NSs for different EoS in a specific alternative theory of gravity. The trained models exhibited high accuracy. Furthermore, this research offers a valuable tool for Bayesian inference methods. The speed-up achieved through the trained ANN models allows for a more efficient exploration of the parameter space.  {\color{black}We plan to construct similar ANN models also for other theories of gravity and for parameterized EoS formulations.}

\subsection{Acknowledgments}

We are grateful to Alexandra Eleni Koloniari, Bhaskar Biswas, and Paraskevi Nousi for their comments.

    \begin{figure*}[t!]
        \centering
        \includegraphics[width=0.8\textwidth]{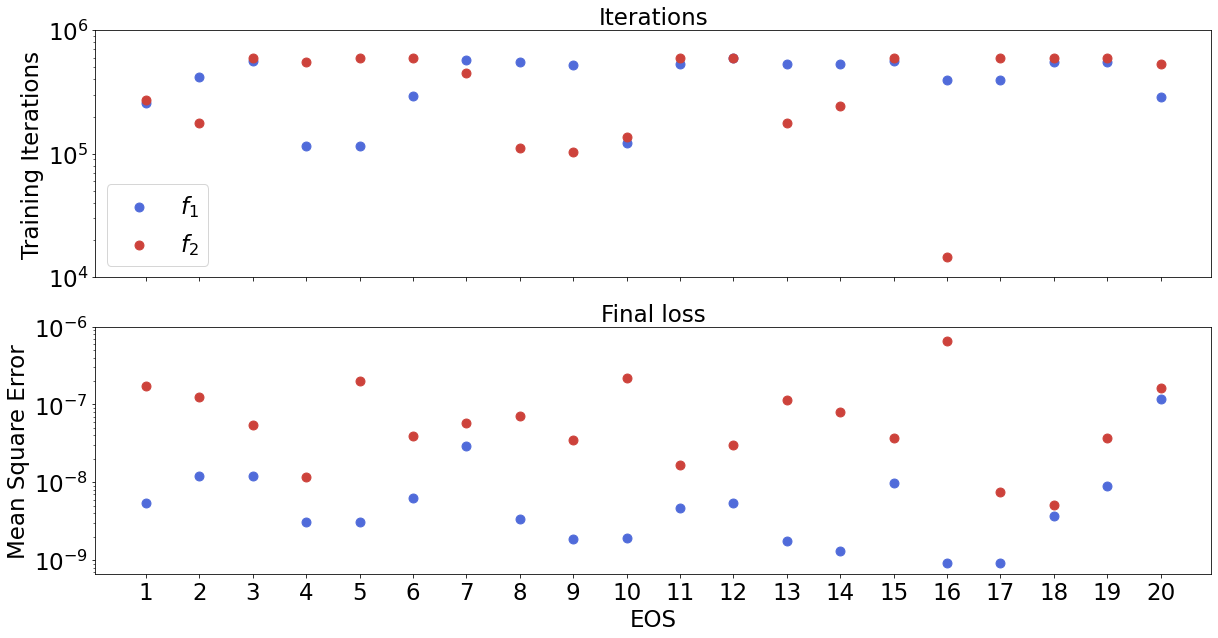}
        \caption{Iterations and final loss for $f_1$ and $f_2$.}
        \label{fig:f1_f2_iters}
    \end{figure*}

\appendix
\section{Training iterations and final loss}
    \label{sec:iters}
        Figure \ref{fig:f1_f2_iters} demonstrates the final loss and the number of iterations for each training. The number of iterations ranges between $10^4$ and $10^6$ while the MSE of the final loss ranges between $8 \cdot 10^{-10}$ and $10^{-6}$. The inconsistency of iterations among training for different EoSs is attributed to the algorithm used for optimization since the number of BFGS iterations is not the same as the number of iterations presented in Figure \ref{fig:f1_f2_iters}. The common variable of every training is the ``maximum iterations", which is the BFGS iterations, and it was set to 230000, not the total iterations. The latter is the quantity of loss-function assessments and will be higher than the maximum iterations since each BFGS iteration must calculate the loss function and the gradients numerous times. There was the case that the algorithm stopped earlier, which indicates that additional stopping conditions were in effect (i.e., the losses did not significantly vary between two iterations).

        One can notice that for every EoS, $f_1$ models reached a lower final loss than $f_2$ models, although based on the dimensionality, the training should be easier for $f_2$. The discrepancy is attributed to the training data difficulty in specific regions for $f_2$ (i.e., for high $M$ values and every $\alpha$ the curves are almost horizontal), which leads to Appendix \ref{sec:spikes}.

       \begin{figure}
            \centering
            \includegraphics[width=0.8\linewidth]{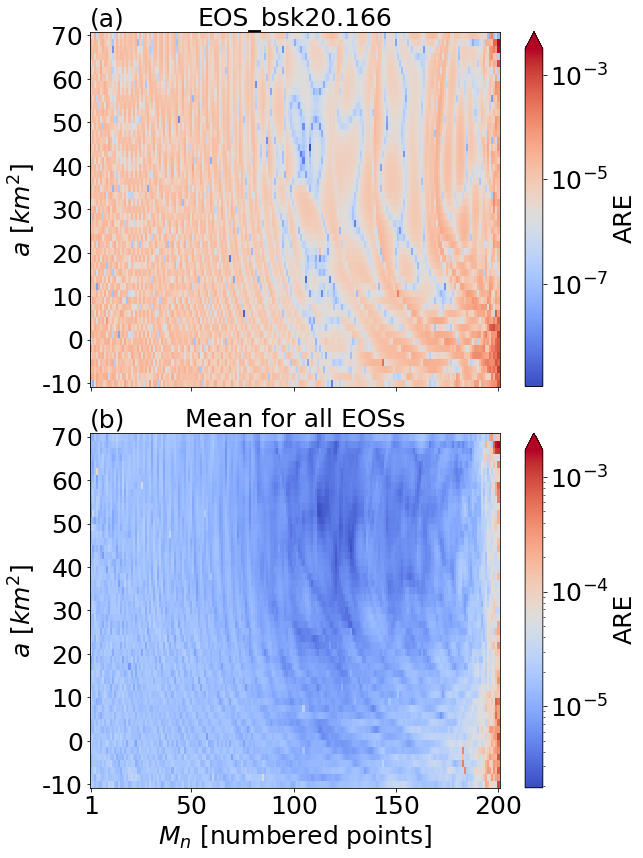}
            \caption{(a) $f_2$ AREs for BSk20 and (b) mean AREs for every EoS. $M_{\rm{n}}$ is the numbered $M$ points with ascending order, ranging from 1-200.}
            \label{fig:pcolorplot_2}
        \end{figure}

                   \begin{figure*}[t!]
            \centering
            \includegraphics[width=0.6\linewidth]{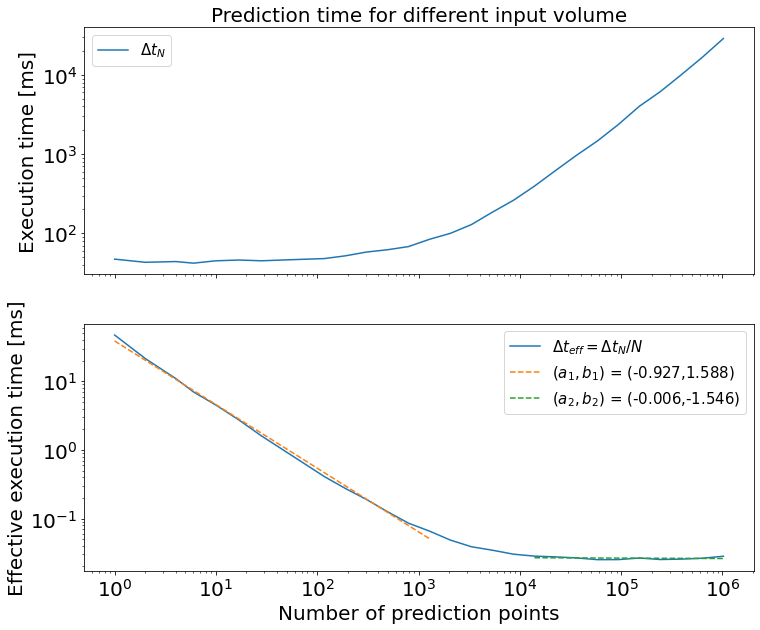}
            \caption{Scaling of the effective run time with the number of points in $\bf X$.}
            \label{fig:pred_time}
        \end{figure*}

\section{Large AREs in $f_2$}
    \label{sec:spikes}
        As mentioned in section \ref{sec:training_res} when discussing Figure \ref{fig:f1_f2_ares}, the Maximum AREs on the testing set for $f_2$ are systematically higher compared to $f_1$, although the MAREs do not demonstrate the same behavior. This is attributed to the nature of $f_2$ data, since for one $\alpha$ and as $M$ increases, the $M-R$ curve reaches an almost horizontal region, making the gradient of $f_2$ very large. This is implied in figure \ref{fig:eos_bsk20_f2}, which, compared to figure \ref{fig:eos_bsk20}, demonstrates larger upper spikes in the testing results. To support this statement, Figure \ref{fig:pcolorplot_2} is presented, where $M_{\rm{n}}$ merely corresponds to the number of $M$ points for every $a = \rm{const}.$ curve, in increasing order. $M_{\rm{n}}$ is used instead of $M$ just to compare the data from every EoS. The top plot of figure \ref{fig:pcolorplot_2} shows the AREs for every input pair and a specific EoS. Here, the AREs are calculated on the whole data set. One can notice that the larger AREs are located at the right edge, corresponding to large $M_{\rm{n}}$ values, while the remaining area seems uniform, except for some local structures of very small AREs. The bottom plot of Figure \ref{fig:pcolorplot_2} shows the mean AREs for all EoSs and for every input pair - which would not be possible if it was not for $M_{\rm{n}}$. Note that these values are not MAREs since MARE refers to mean with respect to data points of the same EoS. Indeed, the mean values demonstrate a uniform behavior around $10^{-5}$, while the right edge is over two orders of magnitude higher. It is, therefore, evident that the largest AREs are located in the near-horizontal region of $M-R$ curves (i.e., for $M_{\rm{n}}$ near 200).

\section{Scaling of the {\tt model.($\bf X$)} run time.}
\label{app:nonlinear}

The top panel in Figure \ref{fig:pred_time} illustrates the total run time {\color{black}as a function of the size of $\bf X$,} ranging from 1 to 1020000 data points\footnote{\color{black}The total run time was measured for 30 different values of the size of $\bf X$ with logarithmic step.}. The bottom panel in Figure \ref{fig:pred_time} presents the effective run time as a function of the size of $\bf X$. We can distinguish two different regions in this plot. One that appears linear in log-log space, ranging from a size of 1 up to a size of 1000 input points, and one that appears nearly horizontal, starting from $10^4$ input points and higher. We calculated a linear fit for each of these regions.

Regarding the first region, we can define $\tilde{y}$ and $\tilde{x}$ as
            $$ \tilde{y} = \log(\Delta t_{eff}) = \log(\Delta t_N) - \log(N), $$
            $$\tilde{x} = \log(N). $$
            With a least-squares fit, we obtain:
            $$ \tilde{y} = a_1\tilde{x} + b_1, $$
            where $a_1 = -0.927$ and $b_1 = 1.588$. Hence:   
            $$ \Delta t_N = 10^{b_1} \times N^{(a_1+1)} \Rightarrow$$
            \begin{equation}
            \label{eq:t_N_1}
                \Delta t_N \left[ ms \right] = 38.711 \times N^{0.073}, \ {\rm for} \ N \in \left[ 1,1000 \right]. 
            \end{equation}
            The final result implies that $\Delta t_N$ and $N$ are related through a power law. Hence, we can select the required total prediction time $\Delta t_N$, and the above Eq. (\ref{eq:t_N_1}) will give the number of points $N$ that we have to insert as input in the model.

            Following the same procedure for the second region, we find
            \begin{equation}
            \label{eq:t_N_2}
                \Delta t_N \left[ ms \right] = 0.028 \times N, \ {\rm for} \ N \in \left[ 10^4, 10^6 \right],
            \end{equation}
            which implies a linear behavior of the total run time for an input size larger than $\sim 10^4$. Eqns (\ref{eq:t_N_1}) and (\ref{eq:t_N_2}) are two branches of the same function and can be used to calculate the expected total run time for a particular number of inputs in the model.

            Method \#2 is intended for consecutive queries of single inputs (i.e., in \textit{for loops}). On the contrary, method \#3 is intended for single queries of a large batch of inputs. Thus, a reasonable question is for which input size it is preferable to use method \#2 in a loop instead of method \#3, in case we do not demand single inputs for other reasons. An approximate calculation is the following. From the data of Figure \ref{fig:f1_ratio}, we can calculate the mean run time of \texttt{model(X)}, which is 1.160 ms. Since \texttt{model(X)} will be in a for loop, we can express the total run time of the loop for $N$ pairs of inputs as
            \begin{equation}
            \label{eq:t_model_X}
                \Delta t_{\rm model(X)} \left[ ms \right] = 1.16 \times N.
            \end{equation}
            For $N < 10^3$ from Eqs. 
            we find
            \begin{equation}
                N = 43.988 \times { \rm ratio}^{1.079}.
            \end{equation}
            Demanding that ${\rm ratio} \geq 1$, we obtain $ N \geq 44$. Thus, for $N$ up to 44, it is faster to use \texttt{model(X)}, while for $N$ greater than 44, it is preferable to use \texttt{model.predict(X)}.

            \bibliography{Liodis-etal}%

\end{document}